\documentclass[a4paper,
]
{book}
\usepackage{epsf_nano}
\usepackage{nano2cmr_arxiv}
\usepackage{textcomp}
\begin{document}

\pnum{}%Paper index, will be inserted later.
\ttitle{Investigation of quantum-dimensional structure parameters by X-ray optical, scanning tunneling and transmission electron microscopy 
}%For Table of Contents. No linebreaks!
\tauthor{ {\em A.~G.~Touryanski}, V.~M.~Senkov, S.~S.~Gizha, L.~V.~Arapkina,  V.~A.~Chapnin, K.~V.~Chizh, V.~P.~Kalinushkin, M.~S.~Storozhevykh, O.~V.~Uvarov, V.~A.~Yuryev}%For Table of Contents.
%No affiliation marks!

\ptitle{Investigation of quantum-dimensional structure parameters by X-ray optical, scanning tunneling and transmission electron microscopy 
}%For Paper title formatting, with linebreaks, if needed.
\pauthor{ {\em \textbf{A.~G.~Touryanski}$^{\,1}$}, V.~M.~Senkov$^{1}$, S.~S.~Gizha$^{1}$,  L.~V.~Arapkina$^{2}$, V.~A.~Chapnin$^{2}$, K.~V.~Chizh$^{2}$, V.~P.~Kalinushkin$^{2}$, M.~S.~Storozhevykh$^{2}$, O.~V.~Uvarov$^{2}$, V.~A.~Yuryev$^{2}$}

\affil{$^{1}$~Lebedev Physical Institute of the Russian Academy of Sciences, 53 Leninskiy Prospekt, Moscow, 119991, Russia
\\ $^{2}$~Prokhorov General Physics Institute of the Russian Academy of Sciences, 38 Vavilov Street, Moscow, 119991, Russia}

\begin{abstract}
{Application of the two-wavelength X-ray reflectometry to exploration of Ge/Si(001) hereostructures with dense  chains of stacked Ge quantum dots is presented.
} 
\end{abstract}

\begindc %Command, starting two columns formatting.

\index{Touryanski A. V.}% For index, put Surname and Initials of each paper author.
\index{Senkov V. M.}   % No special symbols between, as 'nobreakspace' etc.
\index{Gizha S. S.}
\index{Arapkina L. V.}
\index{Chapnin V. A.}
\index{Chizh K. V.}
\index{Kalinushkin V. P.}
\index{Storozhevykh M. S.}
\index{Uvarov O. V.}
\index{Yuryev V. A.}

\section*{Introduction}

For the complete and reliable determination of quantum-dimensional heterostructure parameters it is necessary to use different methods of surface and nanostructure characterization. In this work we successfully used in situ STM, HRTEM and new method: two-wavelength X-ray reflectometry \cite{Touryanski01,Touryanski02}. This X-ray method comprises  measurements of the scattering and reflectivity diagrams at several wavelengths simultaneously for one scan and essentially increases accuracy of identifying parameters  of  investigated objects. It allows to research multilayer structures with diffused  interfaces and permits quantitative analysis of X-Ray reflectometry data down to zero grazing  angle.

\begin{figure}[b]
\leavevmode
\centering{\epsfbox{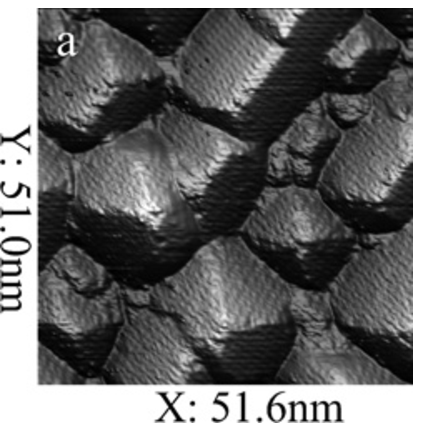}}
\centering{\epsfbox{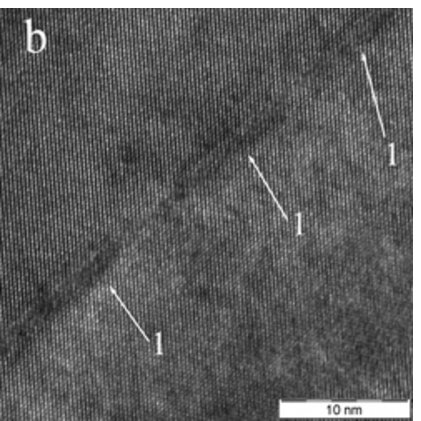}}
% PrincipalAuthor1.eps is figure name in accordance with the file-naming convention.
%\vspace{30mm}% If you do not use epsf-macros, this line put the vertical size of the figure.
\caption{
\label{fig:Tur_STM-HRTEM}
STM (a) and HRTEM (b) images of Ge/Si(001) QDs: $h_{\rm Ge}$ = 10\,\r{A}, $T_{\rm gr}$ = 360{\textcelsius}; in (b), `1' is Ge QDs overgrown by Si.}
\end{figure}

\section{Experimental}

The studied samples were Ge/Si heterostructures with Ge quantum dots (QDs) grown by molecular beam epitaxy (MBE) in Riber EVA32 ultrahigh-vacuum (UHV) MBE chamber at 360{\textcelsius} on Si(001) substrates \cite{Touryanski03, Touryanski04}. On first type of samples, a single layer of uncapped Ge QDs was grown (the Ge coverage $h_{\rm Ge}$ = 10\,\AA). The samples of the second type were Ge/Si heterostructures with five layers of Ge QDs separated by Si spacers with the thickness of 15\,\AA; the cap thickness was 70\,\AA. A study of the heterostructures by STM, carried out by the UHV instrument integrated with the UHV MBE vessel, and by HRTEM has shown the Ge QDs arrays to consist of hut clusters faceted by the \{105\} planes. Ge QDs are seen to occupy whole the sample surface, a free wetting layer surface is not observed (Fig.~\ref{fig:Tur_STM-HRTEM}a). The QDs lateral density amount to  $\sim 5\times 10^{11}$cm$^{-2}$, typical heights of the hut clusters are 10 to 15\,\AA; lenghts of their bases range from 10 to 15\,nm; lengths of wedge-shaped huts reach 30\,nm. Fig.~\ref{fig:Tur_STM-HRTEM}b representing a HRTEM image of Ge QDs located between the Si layers illustrates the internal structure of the studied samples.

X-ray optical measurements were carried out on two-wavelength  arrangement   ``CompleXRay--C6''. The important advantage  of the X-ray optical scheme of this device  comparing with the standard setup is application of  special elements for selection of spectral lines. This X-ray optical scheme is patented in the USA \cite{Touryanski01}. In the present work, the measurements  were performed by means of characteristic lines of copper---CuK$_{\alpha}$ (${\lambda} = 0.154$\,nm) and CuK$_{\beta}$ (${\lambda} = 0.139$\,nm). 

\section{Analysis of Ge/Si heterostructures}

\begin{figure}[b]
% Next two lines used for eps-figure insertion.
%\leavevmode
\centering{\epsfbox{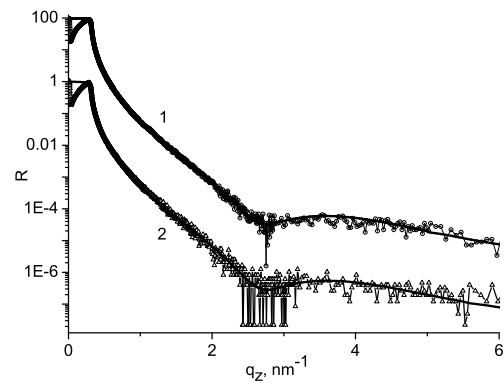}}
% PrincipalAuthor2.eps is figure name in accordance with the file-naming convention.
% If you do not use epsf-macros, this line put the vertical size of the figure.
\caption{
\label{fig:Tur_1-layer}
The X-ray reflectivities of the sample of the first type  for (1)  ${\lambda}=0.154$\,nm and (2) 0.139\,nm; the dots are the experimental data, the solid line is the mathematical simulation; the  curve 2 is  multiplied by 100.}
\end{figure}

The experimental data and the fitting results at two above mentioned wavelengths  for the first type sample (Ge/Si heterostructure with single QDs layer) are presented in Fig.~\ref{fig:Tur_1-layer} 
The results of mathematical simulation---the layer thicknesses $h$ and densities $\rho$---are presented in Table 1.

\begin{table}[h]
\vspace{-12pt}
\caption{}
\begin{center}
\tabcolsep2pt
\begin{tabular}
{lcccccc}
\hline                   
Layer     &    &$h$, nm &    &$\rho$, g/cm$^3$        \\
\hline
GeO$_x$    &       & 0.9  &   & 2.4       \\
\hline
Ge (QDs)    &      & 0.5  &  & 2.5       \\
\hline
Ge (wetting layer) & & 0.4 &    & 3.8      \\
\hline
Si (buffer)      & &   &   & 2.33      \\
\hline
\end{tabular}
\end{center}
\vspace{-12pt}
\end{table}

The results reveal that surface of QDs of the first type sample was oxidized after the sample exposition on air. The general thickness of  layer of oxidized Ge QDs is 1.4\,nm. The average density of this layer is 2.4 to 2.5\,\,g/cm$^3$. The heights of QDs correlate with average value, obtained from STM.
There is a good agreement of experimental and fitted curves both for CuK$_{\alpha}$ and for CuK$_{\beta}$ wave lengths that indicates high accuracy of mathematical simulation of  the structure.

\begin{figure}[t]
% Next two lines used for eps-figure insertion.
%\leavevmode
\centering{\epsfbox{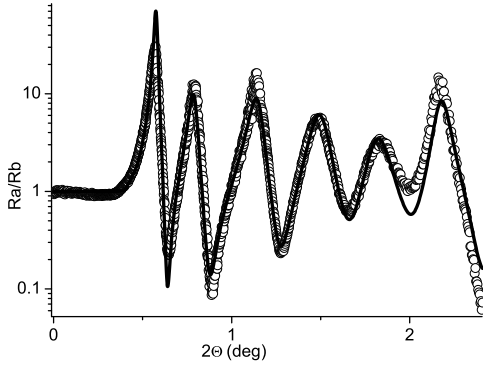}}
% PrincipalAuthor2.eps is figure name in accordance with the file-naming convention.
% If you do not use epsf-macros, this line put the vertical size of the figure.
\caption{
\label{fig:Tur_5-layers}
 Angular dependence of the reflectivity ratio $R_{\alpha}/R_{\beta}$ for the samples of the second type.
}
\end{figure}

The reflectivity angular dependence of the 5-layer  Ge/Si heterostructures  with QDs (the samples of second type) in the relative mode (the reflection coefficients ratio $R_{\alpha}/R_{\beta}$) is presented in Fig.~\ref{fig:Tur_5-layers}. The mathematical simulation of the data of distribution of components through the depth permits to determine the sample structure, that is presented in Table 2. The obtained data coincide well with the STM results.

\begin{table}[ht]
\vspace{-12pt}
\caption{}
\begin{center}
\tabcolsep2pt
\begin{tabular}
{llccllcc}
\hline
{\#}         &Layer     &$h$, nm     &$\rho$, g/cm$^3$ & {\#}          &Layer    &$h$, nm     &$\rho$, g/cm$^3$    \\
\hline
1           & SiO$_2$   &  1.2  &    1.5   &  7           & Ge QDs     & 1.44     & 4.37      \\ 
%\hline
2          & Si (cap)    & 8    & 2.33   & 8          & Si     & 1.84    & 2.33        \\
%\hline
3           & Ge QDs     & 1.27    & 4.5   &  9          & Ge QDs     & 1.21     & 4.13   \\
%\hline
4         & Si     & 1.25     & 2.33   &  10          & Si     & 1.2     &2.33       \\
%\hline
5           & Ge QDs  &  1.03   &   4.11   &   11           & Ge QDs     & 1.93   &  4.28      \\
%\hline
6          & Si     & 1.63    & 2.33 &  12          & Si     &     & 2.33     \\
\hline
\end{tabular}
\end{center}
\vspace{-12pt}
\end{table}

For determination of quantum dot sizes the Grazing Incidence Small Angle X-ray Scattering (GISAXS) in  the plane of incidence was measured. The primary beam was directed to the sample  at  grazing angle $\theta_0=0.236^{\circ}$, that is greater than the critical angle of total reflection for Si ($\theta_c=0.222^{\circ}$ for CuK$_{\alpha}$ and $\theta_c=0.201^{\circ}$ for CuK$_{\beta}$).

\begin{figure}[t]
% Next two lines used for eps-figure insertion.
%\leavevmode
\centering{\epsfbox{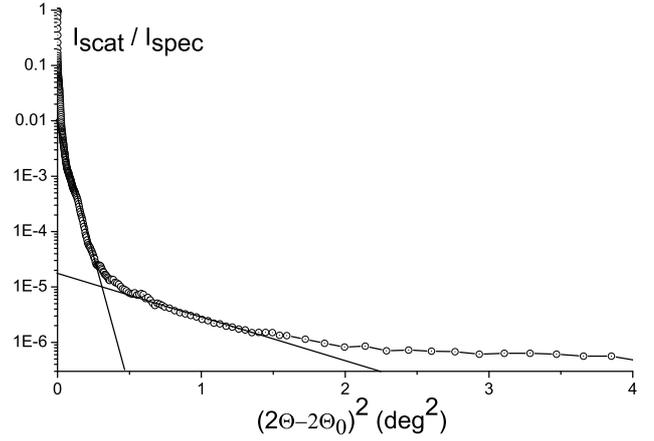}}
% PrincipalAuthor2.eps is figure name in accordance with the file-naming convention.
% If you do not use epsf-macros, this line put the vertical size of the figure.
\caption{
\label{fig:Tur_Gisax}
GISAXS diagram of the sample of the second type; the particle dimensions are 8.5\,nm and 30\,nm.  
}
\end{figure}

The scattering diagram (GISAXS) of the first type sample is given in Fig.~\ref{fig:Tur_Gisax}. The sizes of scattering  centers determined by Guinier relation \cite{Touryanski05} equal 8.5\,nm and 30\,nm under the assumption of the sphericity of particles. These sizes correlate well with  average lateral sizes of  Ge quantum dots.

GISAXS mode enables the determination of cross-correlation in adjacent QDs layers. It is especially important for determination of cross-correlation of QDs in multilayer heterostructures. This factor affects on the photosensitivity, photoluminescence, electrical conductivity, etc.

The scattering diagram (GISAXS) of the second type sample  is presented in Fig.~5. Observed oscillations can be related to position correlation of QDs in structure layers. 

\begin{figure}[t]
% Next two lines used for eps-figure insertion.
\centering{\epsfbox{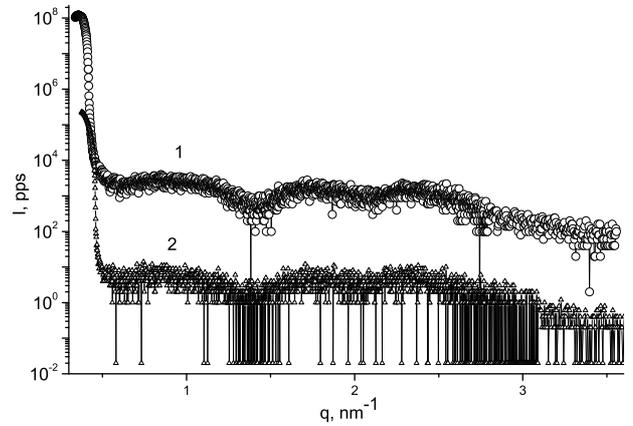}}
\caption{
\label{fig:Tur_sas}
GISAXS diagram of the samples of the second type for (1)  ${\lambda}=0.154$ and (2) 0.139\,nm; 
curve 1 is  multiplied by 100.}

%\leavevmode

% PrincipalAuthor2.eps is figure name in accordance with the file-naming convention.
% If you do not use epsf-macros, this line put the vertical size of the figure.

\end{figure}

\section*{Conclusion}
The obtained results indicate that X-ray reflectometry method can be effectively used for nondestructive, highly informative  researches  of inhomogeneous nanostructures, particularly multilayer heterostructures with QDs arrays. It is also important that X-ray optical methods allow one to obtain a complete information characterizing a whole object.%\\

\acks Equipment of the Center of Collective Use of Scientific Equipment of GPI RAS was used;
MSS was supported by RFBR (grant No.\,12-02-31430${\backslash}$12).

%Journal References

\end{document}